# Towards localized accuracy assessment of remote-sensing derived built-up land layers across the rural-urban continuum

Johannes H. Uhl[1,2], Stefan Leyk[1,2]

[1]Department of Geography, University of Colorado Boulder, Boulder, Colorado, USA
[2]Institute of Behavioral Science, University of Colorado Boulder, Boulder, Colorado, USA

*Corresponding author: Johannes.Uhl@colorado.edu

**Abstract**
The accuracy assessment of remote-sensing derived built-up land data represents a specific case of binary map comparison, where class imbalance varies considerably across rural-urban trajectories. Thus, local accuracy characterization of such datasets requires specific strategies that are robust to low sample sizes and different levels of class imbalance. Herein, we examine the suitability of commonly used spatial agreement measures for their localized accuracy characterization of built-up land layers across the rural-urban continuum, using the Global Human Settlement Layer and a reference database of built-up land derived from cadastral and building footprint data.

With recent technological advances in geospatial data acquisition, processing, as well as cloud-based geospatial data dissemination and analysis infrastructure, there is an increasing amount of novel geospatial datasets available, measuring the spatial(-temporal) distribution of human settlements at large spatial extents and at unprecedented spatial granularity. These datasets include the Global Human Settlement Layer (Pesaresi et al. 2015), Global Urban Footprint (Esch et al. 2013), High-Resolution Settlement Layer (Facebook Connectivity Lab & CIESIN 2016), and the World Settlement Footprint (Marconcini et al. 2019). While such datasets greatly facilitate the study of urbanization, human-natural systems and related geographic-environmental processes at unseen levels of detail, little research has been done on the accuracy of such datasets and how accuracy trajectories can be characterized across the rural-urban continuum, often due to the lack of reliable reference data over sufficiently large spatial extents. Previous work has revealed varying levels of accuracy among different settlement datasets (Klotz et al. 2016), increasing accuracy levels over time in case of the multi-temporal Global Human Settlement Layer (GHSL), and increases in accuracy from rural towards urban areas (Uhl et al. 2017, Leyk et al. 2018, Uhl et al. 2020). However, these general trends are based on coarse, regional stratification of the studied area and density variations derived from the reference data, and thus possibly neglect local accuracy variations.

Several approaches for localized accuracy assessments of categorical spatial data have been proposed in the past (e.g., Leyk and Zimmermann 2004, Foody 2007, Stehman and Wickham 2011), typically applied to (multi-class) land cover data at relative coarse spatial resolutions. High-resolution built-up land data, discriminating between built-up and not built-up land in a binary fashion, exhibit some significant differences with respect to multi-class land cover data:

1) Class imbalance switches between highly urban and sparsely populated areas: The positive class (i.e., "built-up") may be the dominant class in urban areas, but highly underrepresented in rural areas.

2) Localized (i.e., spatially constrained) confusion matrices (of dimension 2x2 in the binary case) characterizing local accuracy may be based on small sample sizes, and possibly contain empty elements, e.g., caused by zero instances of false positives in a local spatial context prohibiting calculation of those measures.

Thus, a framework for localized accuracy assessment of built-up land data needs to account for extreme, bi-directional class imbalance, as well as for low sample sizes underlying a spatially constrained confusion matrix, and the absence of instances of one or more (dis)agreement categories. We are currently developing such a framework, using the GHSL as test data (Figure 1a) and an integrated reference dataset derived from cadastral parcel data and building footprint data (Leyk et al. 2018) as validation data (Figure 1b). We examine the suitability of a variety of commonly used agreement and accuracy measures for local characterization of positional and quantity agreement of built-up land layers, and analyze their interactions with each other, and with variables characterizing the rural-urban continuum. We compute surfaces of Percent Correctly Classified (PCC), User's Accuracy (UA), Producer's Accuracy (PA), Cohen's Kappa, F-measure, G-mean, Intersection-over-union (IoU) for positional agreement, as well as absolute errors (AE) and relative errors (RE) in built-up area as measures of local quantity agreement. Figures 1c,d show such surfaces for PCC and the F-measure, respectively, for Springfield, Massachusetts, USA, illustrating the differences in local accuracy estimated by PCC and the F-measure, particularly in area of lower built-up density, where PCC yields inflated values due to class imbalance (i.e., dominating not-built-up class).

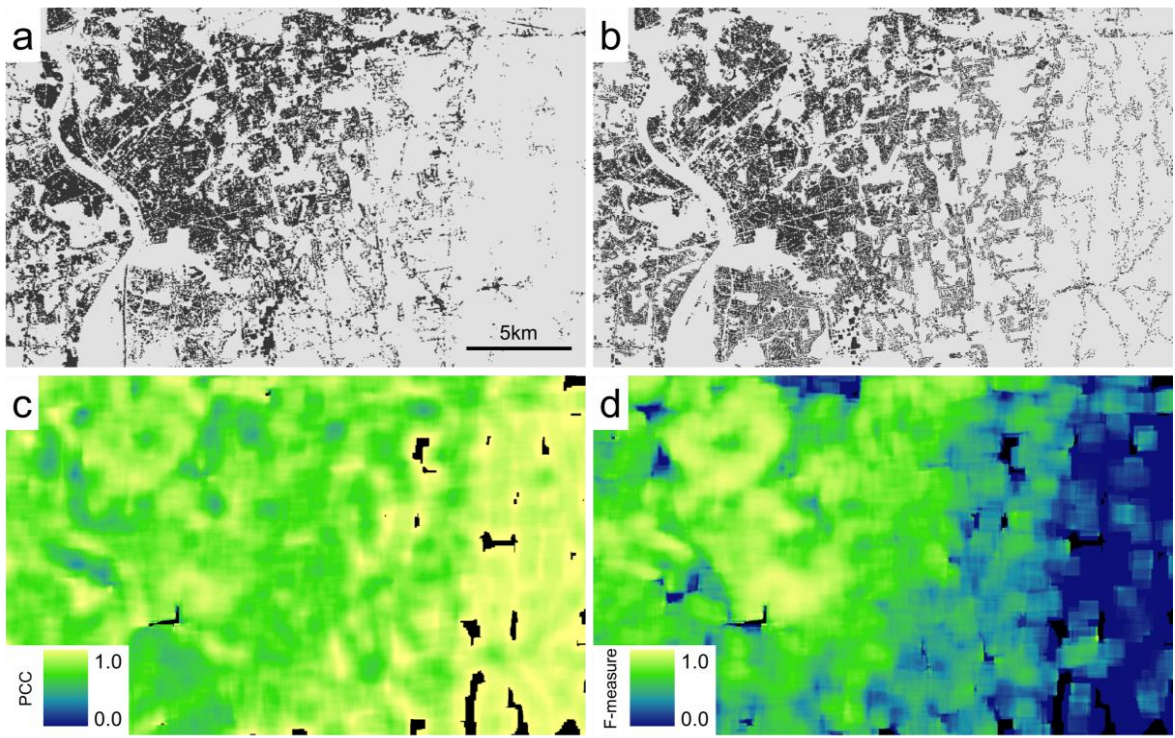

Figure 1: Data examples used in this study, shown for Springfield, Massachusetts, USA: (a) Built-up areas (black) from the GHSL in 2014, (b) reference built-up land surface derived from building footprint and cadastral data, and localized accuracy surfaces for (c) PCC and (d) F-measure, both computed within focal windows of 1x1km. Black areas in (c) and (d) depict no-data areas or areas excluded due to unreliable reference data.

Using the same grid and focal window size, we calculate surfaces of reference built-up area density, and of focal landscape metrics (e.g, the area of the largest built-up patch), derived from the reference data, to characterize the density and structure of built-up areas. We assume such metrics to vary with the rural-urban gradient and thus, representing a proxy measure of the rural-urban continuum, ranging from scattered, sparse rural settlements to dense, and highly connected built-up areas in urban settlements.

Figure 2 shows the relationships between these metrics and local measures of built-up density and structure, computed for the state of Massachusetts and GHSL built-up labels in 2015, revealing interesting, partially contradicting trends of the tested measures across the rural-urban continuum. Moreover, the variability in these scatterplots indicates high levels of variation among measures due to the conservativeness in their mathematical structure. For example, Kappa index never exceeds the F-measure, and achieves similar values in regions of low built-up density, and where the largest built-up patch area is low (i.e., areas of sparse, scattered settlements).

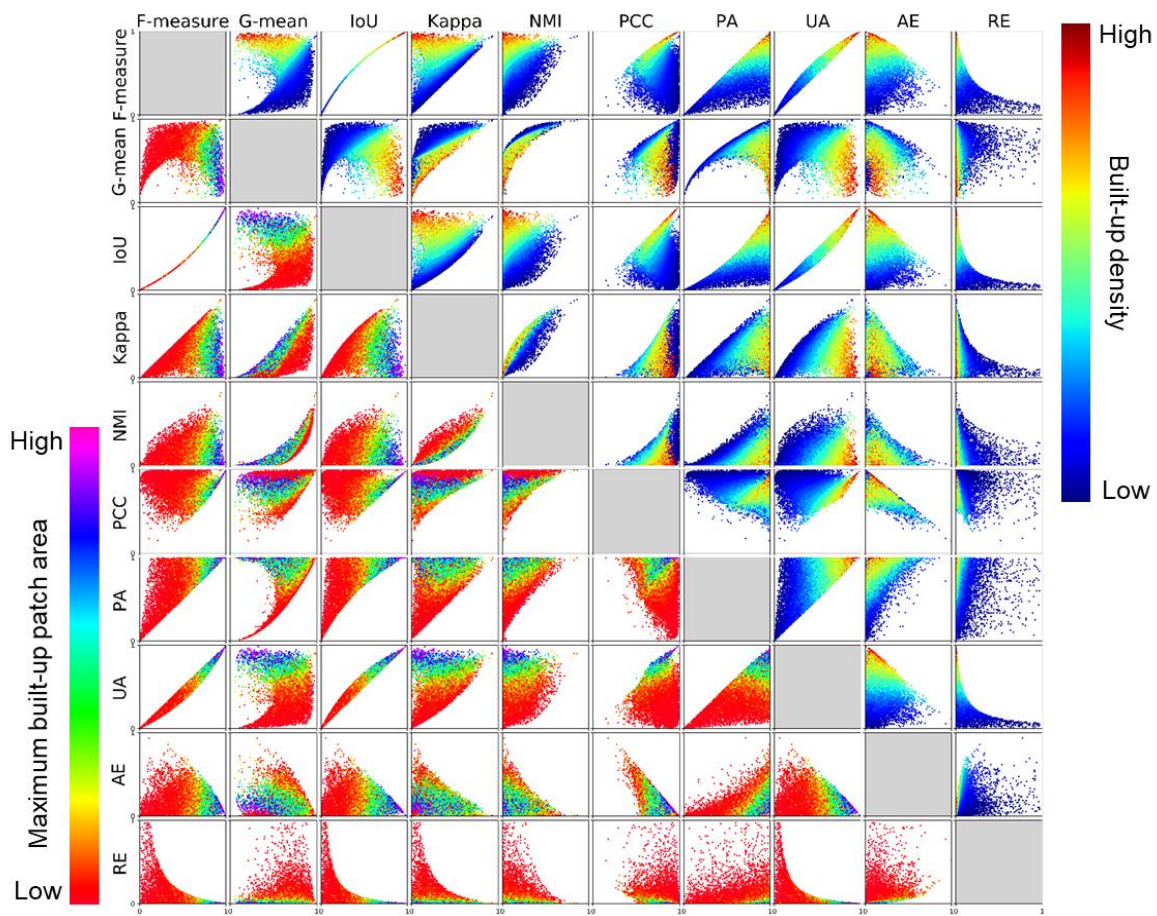

Figure 2: Scatterplot matrix generated by pixel-wise comparison of localized positional and quantity agreement surfaces computed at 30m spatial resolution and within 1x1km focal windows for the state of Massachusetts, USA.

Future work includes the analysis of effects of varying spatial support (i.e., the size of focal windows used to generate localized confusion matrices), and of the analytical unit on local accuracy characterization, as well as the suitability of built environment and socio-economic variables for uncertainty prediction of remote-sensing derived built-up land data.

**ACKNOWLEDGMENT**

Support for this work was provided through the Eunice Kennedy Shriver National Institute of Child Health & Human Development of the National Institutes of Health under Award Number P2CHD066613. The content is solely the responsibility of the authors and does not necessarily represent the official views of the National Institutes of Health.

**REFERENCES**

Esch, T., Marconcini, M., Felbier, A., Roth, A., Heldens, W., Huber, M., Schwinger, M., Taubenböck, H., Müller, A. and Dech, S.J.I.G., (2013). Urban footprint processor—Fully automated processing chain generating settlement masks from global data of the TanDEM-X mission. *IEEE Geoscience and Remote Sensing Letters*, *10*(6), pp.1617-1621.

Facebook Connectivity Lab and Center for International Earth Science Information Network - CIESIN - Columbia University, (2016). High Resolution Settlement Layer (HRSL). Source Imagery for HRSL © 2016 DigitalGlobe Available at https://ciesin.columbia.edu/data/hrsl/. (Accessed 23-03-2018)

Foody, G.M. (2007). Local characterization of thematic classification accuracy through spatially constrained confusion matrices. *International Journal of Remote Sensing,* 26(6), 1217-1228.

Klotz, M., Kemper, T., Geiß, C., Esch, T. and Taubenböck, H., (2016). How good is the map? A multi-scale cross-comparison framework for global settlement layers: Evidence from Central Europe. *Remote Sensing of Environment*, *178*, pp.191-212.

Leyk, S., Uhl, J.H., Balk, D. and Jones, B., (2018). Assessing the accuracy of multi-temporal built-up land layers across rural-urban trajectories in the United States. *Remote sensing of environment*, *204*, pp.898-917.

Leyk, S., and Zimmermann, N.E. (2004). A predictive uncertainty model for field-based survey maps using generalized linear models. *International Conference on Geographic Information Science,* 191-205.

Marconcini, M., Metz-Marconcini, A., Üreyen, S., Palacios-Lopez, D., Hanke, W., Bachofer, F., Zeidler, J., Esch, T., Gorelick, N., Kakarla, A. and Strano, E., (2019). Outlining where humans live--The World Settlement Footprint 2015. *arXiv preprint arXiv:1910.12707*.

Pesaresi, M., Ehrlich, D., Ferri, S., Florczyk, A., Freire, S., Haag, F., Halkia, M., Julea, A.M., Kemper, T. and Soille, P., (2015). Global human settlement analysis for disaster risk reduction. *The International Archives of Photogrammetry, Remote Sensing and Spatial Information Sciences*, *40*(7), p.837.

Stehman, S.V., and Wickham, J.D. (2011). Pixels, blocks of pixels, and polygons: Choosing a spatial unit for thematic accuracy assessment. *Remote Sensing of Environment,* 115(12), 3044-3055.

Uhl, J.H. and Leyk, S., (2017). Multi-Scale Effects and Sensitivities in Built-up Land Data Accuracy Assessments. *Proceedings of International Cartographic Conference 2017,* Washington D.C., USA. 2017.

Uhl, J.H., Zoraghein, H., Leyk, S., Balk, D., Corbane, C., Syrris, V. and Florczyk, A.J., (2020). Exposing the urban continuum: Implications and cross-comparison from an interdisciplinary perspective. *International Journal of Digital Earth*, *13*(1), pp.22-44.